\documentclass[sigconf,natbib=true]{acmart}

\usepackage{hyperref}
\usepackage{wrapfig,lipsum,booktabs}

\AtBeginDocument{%
  }

\begin{document}

\title{The Presence and the State-of-Practice of Software Architecting in the Brazilian Industry: Preliminary Results of A Survey}

\author{Valdemar Vicente Graciano Neto}
\affiliation{
  \institution{Federal University of Goiás}
  \city{Goiânia}
  \state{GO}
  \country{Brazil}
}
\email{valdemarneto@ufg.br}

\author{Mohamad Kassab}
\affiliation{
  \institution{Boston University}
  \city{Boston}
  %\state{GO}
  \country{USA}
}
\email{mkassab@bu.edu}

\author{Diana Lorena Santos}
\affiliation{
  \institution{Federal University of Goiás}
  \city{Goiânia}
  \state{GO}
  \country{Brazil}
}
\email{dilorena@discente.ufg.br}

\author{Andrey Gonçalves França}
\affiliation{
  \institution{Amazon Brazil}
  %\city{Goiânia}
  %\state{GO}
  \country{Brazil}
}
\email{andreygfranca@gmail.com}

\author{Edson OliveiraJr}
\affiliation{
  \institution{State University of Maringá}
  \city{Maringá}
  \state{PR}
  \country{Brazil}
}
\email{edson@din.uem.br}

\author{Rafael Z. Frantz}
\affiliation{
  \institution{Northwestern Regional University of the State of Rio Grande do Sul}
  \city{Ijuí}
  \state{RS}
  \country{Brazil}
}
\email{rzfrantz@unijui.edu.br}

\author{Ahmad Mohsin}
\affiliation{
  \institution{Edith Cowen University}
  \city{Joondalup}
  %\state{PR}
  \country{Australia}
}
\email{a.mohsin@ecu.edu.au}

\author{Marcos Kalinowski}
\affiliation{
  \institution{Pontifical Catholic University of Rio de Janeiro (PUC-Rio)}
  \city{Rio de Janeiro}
  \state{RJ}
  \country{Brazil}
}
\email{kalinowski@inf.puc-rio.br}

\renewcommand{\shortauthors}{Graciano Neto et al.}

\acmConference[2025]{}{2025}{Brazil}

%\maketitle

\begin{abstract}
\textbf{Context: } Software architecture profoundly influences the overall quality of software systems. Consequently, professionals responsible for designing, maintaining, and evolving software architectures must possess the requisite knowledge and skills to avoid compromising product quality.
\textbf{Objective: } This study aims to characterize how companies in Brazil incorporate the activity of software architecting, examining whether such dedicated professionals are formally employed and, if not, how their responsibilities are redistributed.
\textbf{Method: } The research adopts a survey-based approach to gather evidence from professionals who assume the role of software architects (either formally or not). Descriptive statistics and thematic analysis were then applied to interpret the collected data.
\textbf{Results: } Data were obtained from 105 professionals across 24 Brazilian states. The findings indicate that: (i) not all companies employ a dedicated software architect; (ii) in some cases, other professionals assume architectural responsibilities; and (iii) even in organizations with a formally designated architect, other roles may still perform architect-level tasks.
\textbf{Conclusions: } Professionals explicitly hired as software architects tend to receive higher salaries compared to those carrying out similar architectural activities under different job titles. Nonetheless, many non-architect professionals continue to engage in core architectural duties, reflecting ongoing variations in the formalization and recognition of the architect role. \end{abstract}


\maketitle

\section{Introduction}
\label{INTRODUCTION}

Software architecture can ensure quality by means of a set of design decisions that come prior to the software detailed design. It can assure scalability, availability, security and other important quality attributes (QAs) for modern systems \cite{Pantoja_et_al2024}. %Software architects oversee system design, make key decisions, and maintain consistency across technology stacks \cite{Hohpe_et_al2016}.
A well-designed software architecture demands dedicated architects, who can provide a comprehensive view of software integrity \cite{fowler:2003}. They bridge technical and business teams, translating needs into technical requirements and reducing misalignment risks. %Their ability to code alongside developers and explain technical consequences makes them indispensable. 
Without them, companies risk overloaded teams and declining software quality, potentially harming business outcomes. 

Despite clear benefits, many Brazilian companies lack dedicated architects or delegate responsibilities to other IT roles, threatening software quality. This is concerning as Brazil's software market is projected to reach \$12.57 billion by 2029, growing at 5.98\% annually \cite{statista}. As software becomes a strategic business tool, skilled architects are increasingly crucial. According to a report by Gartner \cite{burke2020top}, organizations are experiencing a surge in demand for skilled software architects to manage complex, distributed systems. In emerging economies similar to Brazil, such as India and China, concerted efforts have been made to formalize software architecture roles to enhance competitiveness and ensure the delivery of reliable, high-quality software solutions. However, little research has explored how Brazilian companies integrate this role, current practices, or challenges in developing this profession. The absence of knowledge about how software architects play their role in Brazilian companies can also hamper a proper feedback loop between the training and education of architects (whether through university programs or specialized certification and practice courses) and the practices currently adopted by the industry.

This study addresses the gap in understanding how software architecture is managed in Brazilian companies by investigating current practices. It aims to provide insights into the role of software architects and help improve development processes. A survey of 105 professionals across 24 of Brazil’s 27 states collected data on demographics, company profiles, responsibilities, and compensation. The questionnaire, divided into four sections, provided a diverse snapshot of software architecture practices. Using descriptive statistics and thematic analysis, we present a comprehensive view of the state of software architecture in Brazil.

%The remainder of this paper is structured as follows: Section \ref{BACKGROUND} offers a comprehensive background review of the role of software architects, covering key responsibilities and practices in both Brazilian and international contexts; Section \ref{survey} details the survey methodology, outlining the design, execution, and data analysis procedures; Section \ref{sec:reporting} presents the study's findings, followed by Section \ref{discussion}, which provides a thorough analysis of these results and their broader implications; Section \ref{threats} addresses potential threats to the validity of the research; and Section \ref{final} concludes with key insights and recommendations for future research.

%\vspace{-0.5cm}

\section{The Role of the Software Architect}
\label{BACKGROUND}

Software architecture involves designing a system’s core elements, their relationships, and guiding principles that shape its evolution \cite{Bass:2012}. Beyond structure, it embodies a shared understanding among developers, ensuring coherence in system design \cite{fowler:2003}. Effective architecture serves as both a technical blueprint and a communication medium. Software architects tackle technical and organizational challenges. Said et al. \cite{said2025modulith} highlight modular architecture’s benefits, while stability depends on architects’ expertise. Another study \cite{rehman2018roles} links architects’ coding roles to reduced bugs and improved outcomes. Findings in \cite{bany2024influence} stress architects’ impact on team collaboration and project success. Their role extends beyond design to selecting technologies and ensuring alignment with architectural vision \cite{Bass:2012}, requiring deep technical and domain knowledge \cite{KRUCHTEN20082413}.
%Architects also manage teams and align projects with business goals \cite{Bass:2012}. They mentor developers, facilitate stakeholder communication, and foster collaboration to ensure architectural decisions support strategic objectives \cite{KRUCHTEN20082413}.  
\\\\
\noindent\textbf{Related Work.} There is limited research on software architects' roles in the Brazilian industry. Santos and Ito \cite{correlato1} examined São Paulo’s architectural origins through an inductive survey of 36 IT professionals, offering localized insights. Research in \cite{figueiredo2014knowledge} used case studies to analyze architects’ role in knowledge adaptation.

Globally, surveys on software architects often focus on specific regions. In Chile, Pérez et al. \cite{perez2019familiarity} studied architects’ perceptions of technical debt, with similar research in Colombia \cite{Perez2020}. %Both found strong awareness of technical debt's causes, emphasizing better management. 
Multinational studies \cite{PEREZ2021, WAN2023} provided broader insights but did not account for Brazil’s industry, often sampling companies where architects are already well established.

Our research expands the scope by including professionals performing architectural duties without formal titles. Surveying 105 professionals across 24 states, we examined role prevalence, responsibilities, and challenges in Brazilian software architecture. Unlike prior studies, our approach ensures national coverage, addressing a critical gap and providing insights into the evolving role of software architects in Brazil.

\section{Research Method}
\label{survey}

The research method is based on a structured ``Questionnaire Survey'' approach, which aligns with established protocols and guidelines used in prior survey-based studies and guidelines in the field \cite{kitchenham2002preliminary,linaker,molleri2016,kasunic2018,Lebtag2021,David2023}.

\subsection{Protocol and Planning}
\label{sec:protocol}

This study seeks to investigate the presence and role of software architects within Brazilian companies. The specific objectives include examining the performance of software architects in Brazil, identifying other professional roles involved in software architecture, evaluating average salary ranges, and mapping the activities undertaken by these professionals in their respective organizations. 

Given the objectives, the following research questions (RQ) were raised:

\textbf{\textit{RQ1 - What are the characteristics of companies involved in software development or maintenance projects in Brazil, either with or without software architects?}}. This question seeks to gather detailed information about the companies where survey respondents are employed. It aims to capture aspects such as company size (based on employee count), location within Brazil, whether the company operates as a multinational, and whether software development is a core business activity or a supporting function.

\textbf{\textit{RQ2 - What is the prevalence and role of software architects in Brazilian companies?}}. The purpose of this question is to explore the presence of software architects in these companies, including whether they are formally hired as software architects or informally perform architecture-related tasks. It also aims to collect data on compensation levels for these roles.

\textbf{\textit{RQ3 - What is the current state of software architecture practices in companies in Brazil?}}. This question aims to understand the specific responsibilities and practices of software architects in Brazilian companies, shedding light on the day-to-day functions and the maturity of software architecture processes within these organizations.

\subsubsection{Questionnaire Design}

The questionnaire was designed to align with the research objectives and target population. It consisted of predominantly closed-ended questions aimed at systematically capturing respondents' demographic data, company profiles, and details about the roles and responsibilities of software architects. The instrument was developed based on best practices in survey research and software engineering, and it underwent a pilot test with three respondents: one PhD with expertise in software engineering and two architects from industry. Following the pilot, minor revisions were made to improve clarity and relevance (e.g., modifying educational level options and salary ranges).

The final version of the questionnaire contained four sections: \textbf{Section 1}: Informed consent and study introduction; \textbf{Section 2}:  Demographic information of respondents (e.g., age, gender, educational background); \textbf{Section 3}: Company profiles (e.g., size, sector, software development as core or support activity); and \textbf{Section 4}: Roles and practices of software architects (e.g., main activities, salary, number of architects per project). The list of questions is provided in Table \ref{tab:demograficas}.

\begin{table}%{l}{6cm}

%Without H, it is not working. Let it there as is.
%\begin{table}[!h]
    \centering
    \caption{Extraction Form and Correspondences to RQs. \textbf{Question Types}: \textit{Open Question (Text-based)} = OQ; \textit{Closed Question (Multiple Choice) = CQ.}}
    \label{tab:demograficas}
    \resizebox{0.45\textwidth}{!}{ 
    \begin{tabular}{|c|p{10cm}|c|}
    \hline \textbf{ID} & \textbf{Questions and Options (for closed questions)} & RQ\\
    \hline OQ01 & How old are you? & - \\
    \hline CQ01 & What is your gender? (Male, Female, I prefer not say it) & -  \\
    \hline OQ02 & What is your city and state? & - \\
    \hline CQ02 & What is your education level? (Elementary School, High School, Undergraduate Studies in Progress, Undergraduate Degree, 
Postgraduate Lato Sensu (Specialization, MBA), Master's Degree, Doctorate (Ph.D.)) & - \\
    \hline  & \textbf{Questions about the Company} & \\
    \hline OQ03 & What is the company name? & RQ1 \\
    \hline
    CQ03 & Is software production an end activity? i.e, is a software production or other business, but who develops software for internal consumption? (Software development is a core activity; It is a company from another sector but develops software for internal use.) & RQ1\\
    \hline CQ04 & What is the company's field of activity? (Software Production, Military, Financial, Educational, Governmental, Others)& RQ1 \\
    \hline CQ05 & Is the company multinational? (Yes, No) & RQ1 \\
    \hline CQ06 & Does the company have branches? (Yes, No) & RQ1\\
    \hline CQ07 & If yes, do you work at the head office or at a branch? (Headquarters, Branch, Not applicable) &  RQ1\\
    \hline CQ08 & What is the number of employees in the unit where you work? (0-10; 11-30; 31-50; 51-100; 101-1000, More than 1000) & RQ1\\
    \hline CQ09 & What is the size of the company? (Microenterprise, Small-sized Company, Medium-sized Company, Large-sized Company)  & RQ1\\
    \hline OQ04 & How many years has the company been in the market? & RQ1\\
    \hline OQ05 & How many customers do you have on average? & RQ1\\
    \hline OQ06 & How many projects do you have in progress? & RQ1\\
    \hline CQ10 & What is the company's sector? (Public, Private) & RQ1\\
    \hline  & \textbf{Questions about the Role of the Architect} & \\
    \hline
    CQ11 & Has anyone in the company been hired specifically to fill the software architect position?  (Yes, No) & RQ2\\
    \hline OQ07 & If so, what are the main activities performed by this professional? & RQ3\\
    \hline OQ08 & If yes, how many software architects are there in your company? & RQ2 \\
    \hline
    {CQ12} & There is someone in the company who performs software architect activities but who does not exercise this function formally? That is, he is a programmer,  manager or other role, but perform architecture activities? (Yes, No) & RQ3\\
    \hline OQ09 & If yes, how many professionals fit in this case? & RQ2 \\
    \hline CQ14 & What is the company's software architect monthly salary range? (Up to R\$ 2,500.00, Between R\$ 2,501.00 and R\$ 4,000.00, Between R\$ 4,001.00 and R\$ 8,000.00, Between R\$ 8,000.00 and R\$ 10,000.00, Between R\$ 10,000.00 and R\$ 20,000.00, Above R\$ 20,000.00, Not Applicable) & RQ2\\
    \hline
    CQ15 & If there is someone who is not an architect but acts as an architect, what is the salary of that professional? (Up to R\$ 2,500.00, Between R\$ 2,501.00 and R\$ 4,000.00, Between R\$ 4,001.00 and R\$ 8,000.00, Between R\$ 8,000.00 and R\$ 10,000.00, Between R\$ 10,000.00 and R\$ 20,000.00, Above R\$ 20,000.00, Not Applicable) & RQ2\\
    \hline OQ10 & How long have professionals in your company been working as architects? & RQ3\\
    \hline OQ11 & How many architects are there per project? & RQ3\\
    \hline CQ13 & Is there an architect who works on more than one project? (Yes, No, Not Applicable) & RQ3\\
    \hline
\end{tabular}
}
\end{table}
%\end{wraptable} 

\subsubsection{Target Population and Sampling Strategy} 

The target population for this study included professionals who assume the role of software architects (either formally or not) in  the industries in Brazil. %We aimed to capture the practices of both companies with formally hired software architects and those where other professionals perform similar roles without the formal title. 
The sampling strategy utilized convenience and snowball sampling, where initial participants were invited via Slack, LinkedIn, WhatsApp, and Telegram groups. Participants were encouraged to share the survey with peers, broadening the reach of the study.

\subsection{Data Collection and Monitoring}

The questionnaire was distributed online. Invitations were sent via online IT communities, targeted emails, and shared links on professional networking platforms. Participation was voluntary, and no financial incentives were offered. %To minimize non-responses and ensure data integrity, regular monitoring was conducted. 
Responses were checked for duplicates, and any incomplete or conflicting entries were discarded. Of the 111 initial responses, 105 were deemed valid for analysis. Participants who did not respond to all questions had their missing data handled through listwise deletion, ensuring that only complete data sets were included in the analysis. The final set of responses represents a diverse range of company sizes, sectors, and regions across 24 of the 27 Brazilian states. Respondents varied in terms of age, gender, educational background, and years of professional experience, ensuring a broad representation of the software architecture profession in the country.

\subsection{Data Analysis}

The collected data was analyzed using a combination of descriptive statistics and thematic analysis for open-ended responses. Closed-ended questions were quantified, providing insights into the prevalence of software architects, their roles, and salary ranges across different companies.  We utilized thematic analysis following the analytical protocol outlined by \cite{Alami2020}. Thematic analysis, a method for identifying, analyzing, and interpreting patterns within qualitative data, offers a flexible yet powerful tool for deriving data about the phenomenon under investigation. %This approach is predominantly inductive, meaning 
The themes emerge organically from the data rather than being predefined before data collection and analysis \cite{flick2022}.

Data underwent a coding process, wherein the text was systematically scrutinized to extract meaningful segments. These segments were subsequently labeled according to their fundamental thematic ideas, facilitating the organization and classification of the data. %Through this coding procedure, a conceptual model was developed, providing a comprehensive explanation of the activities performed by software architects.

\section{Communication of Results}
\label{sec:reporting}

\subsection{Demographics}

\begin{figure*}[!h]
%\begin{wrapfigure}{l}{0.3\textwidth}
  \centering
   \includegraphics[width=0.80\textwidth]{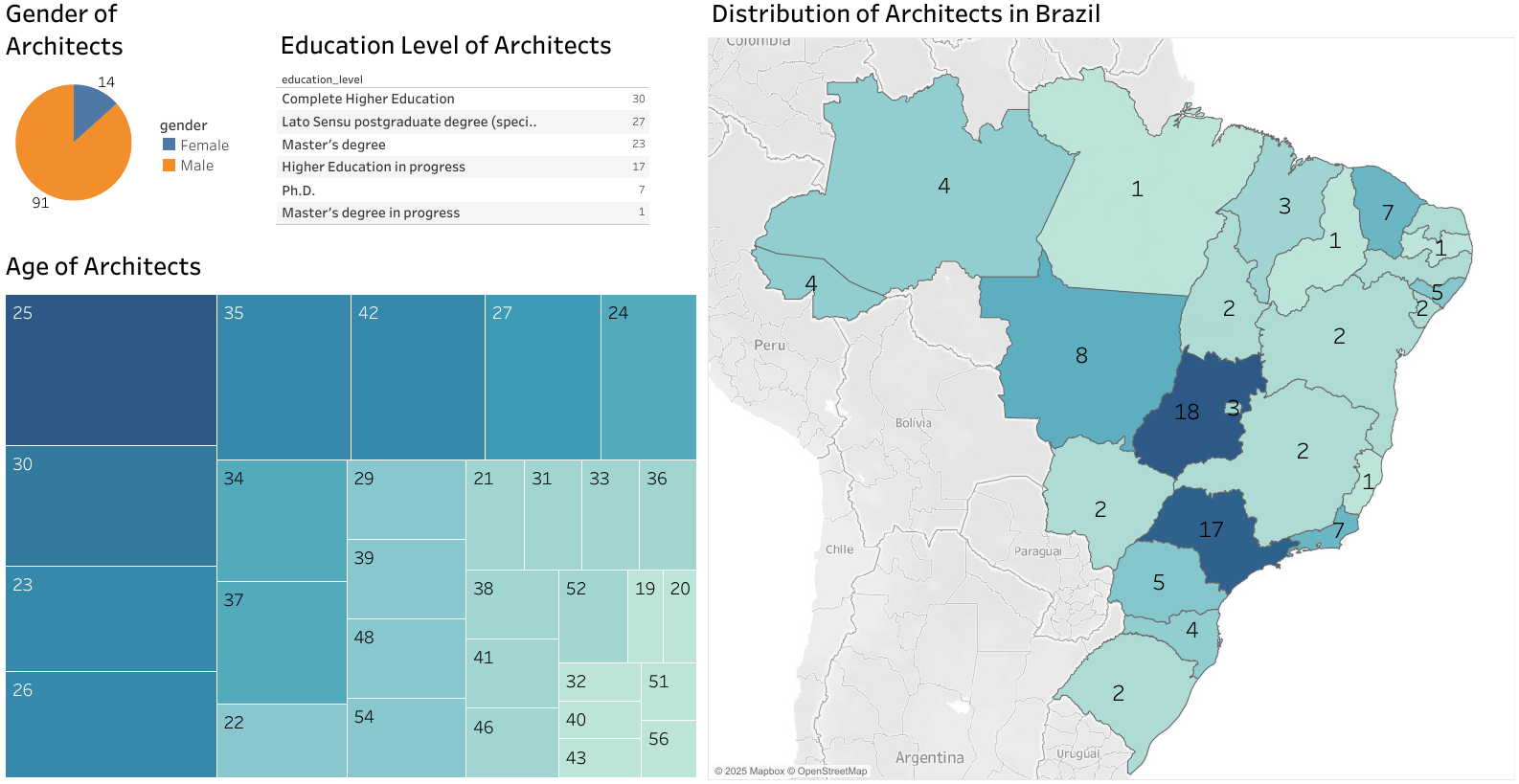}
  \caption{Survey Demographics}
  \label{fig:demographics}
%\end{wrapfigure}
\end{figure*}

The dataset collected via the Google Form is publicly accessible\footnote{\url{https://docs.google.com/spreadsheets/d/1cbKHUZgcSC0Ivr5z4FyeFeYE3_jqZJmWAzSieNBx3ZM/edit?usp=sharing}}. Out of the 105 valid responses received, the average age of the respondents was 32.79 years, with a median age of 30 years. The age range spanned from the youngest respondent at 19 years to the oldest at 56 years (Fig. \ref{fig:demographics}). A majority of the respondents, 91 individuals (86.67\%), identified as male, while 14 (13.33\%) identified as female, as it can be seen in Fig. \ref{fig:demographics}. In terms of educational qualifications, 30 respondents (28.57\%) had completed higher education, 27 (25.71\%) held a Lato Sensu postgraduate degree (specialization or MBA), 23 (21.91\%) had obtained a Master's degree, 17 (16.19\%) were pursuing an undergraduate degree, seven (6.67\%) held a PhD, and one respondent (0.95\%) was enrolled in a Master's program. 70 different companies were identified in the data. 75 of the respondents (71.42\%) provided their company's name (5 were from one of the companies between the 70 different ones) and 30 respondents (28.57\%) preferred not answering that.

The survey elicited responses from professionals residing in 37 cities across Brazil, 22 of which are state capitals. However, no responses were received from the capitals Porto Velho (RO), Boa Vista (RR), João Pessoa (PB), Belo Horizonte (MG), or Macapá (AP). Of the 27 Brazilian states (26 states and the Federal District), responses were obtained from 24 states, with no representation from Amapá, Rondônia, or Roraima. Figure \ref{fig:demographics} depicts such results.

\begin{figure*}[!h]
  \centering
   \includegraphics[width=1\textwidth]{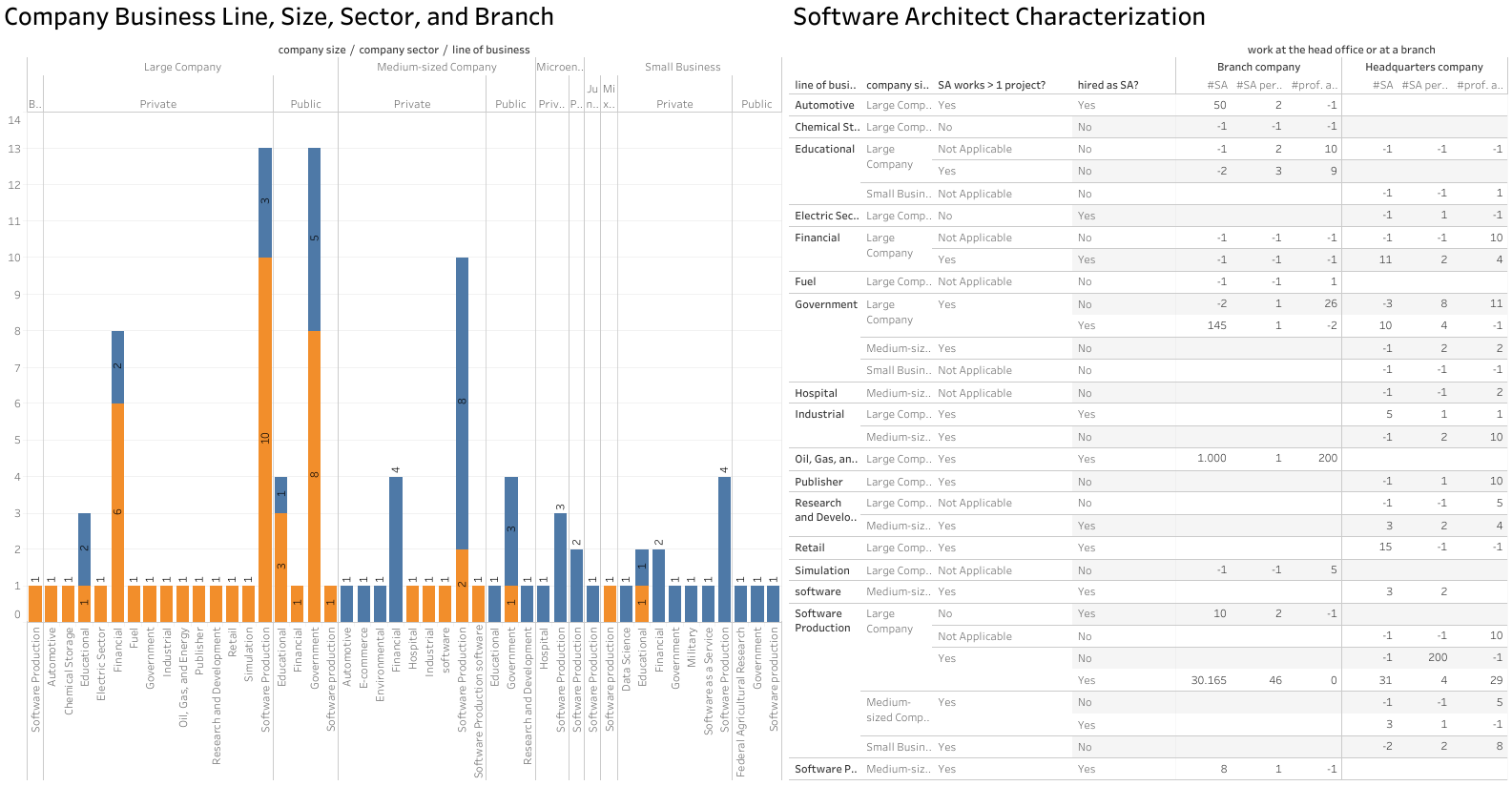}
  \caption{Companies and Software Architects Characterization}
  \label{fig:company_SAs}
\end{figure*}

The following sections present results based on the defined RQs, supported by Figure \ref{fig:company_SAs}. Note that ``-1'' and ``Not Applicable'' mean the respondent could not inform such data.

\subsection{RQ1: Characteristics of Companies}

Research Question 1 (RQ1) focused on the characteristics of companies employing the respondents. Of the 105 respondents, 60\% indicated that their companies’ main business was software production, while 39\% worked for companies in other sectors that develop software for internal use. One respondent represented a company primarily focused on IT service consumption.
\\\\
\noindent\textbf{Company Sectors:} The majority of respondents (37\%) worked in the software sector. Government (19\%) and finance (14\%) followed, while education accounted for 10\%. Healthcare, industrial, and R\&D sectors each comprised 2\%, with niche sectors like energy, automotive, and e-commerce together making up 13\%.

\noindent\textbf{Company Broadening:} Most respondents (74\%) worked for domestic companies, while 26\% were employed by multinationals. Roughly half of the companies (49\%) had branches. Of those, most respondents worked at headquarters, with fewer at branch locations.

\noindent\textbf{Number of Employees:} The largest group of respondents (35\%) worked for companies with 101–1,000 employees. About 21\% were in larger firms with over 1,000 employees, while 16\% were in smaller firms with 11–30 employees. The remainder worked in companies of varying sizes. 

\noindent\textbf{Company Age and Customer Base:} Most companies (37\%) were between 11 and 30 years old, while 26\% were younger than 10 years. One company had been operating for over 500 years. Regarding customer base, 22\% of respondents worked for companies serving 1–10 customers, and 16\% each worked for firms serving 11–100 or 101–15,000 customers. Around 17\% reported working for companies with over 1 million customers.

\noindent\textbf{Projects in Parallel:} Most companies (44\%) handled up to 10 concurrent projects, with 24\% managing 11–50 projects at a time. A small percentage (9\%) managed over 1,000 projects, while 11\% provided inconclusive data.

\subsection{RQ2: Presence of architects}
\label{subsec:Q02}

Research Question 2 (RQ2) investigates the role and presence of software architects in companies, their activities, and their compensation structures.

\noindent\textbf{Presence of Software Architects:} Among the 105 respondents, 59\% indicated that their companies do not have a dedicated software architect, while 41\% confirmed the presence of such a role. However, 68\% mentioned that someone without the formal title of software architect still performs the role, whereas 32\% stated no one in their organization undertakes this function.

When asked about the number of architects in their companies, 27\% reported having up to 10 architects, 6\% had between 11 and 50, and only 3\% had more than 50. Notably, 63\% of the respondents either provided inconclusive answers or did not know. In terms of architects per project, 42\% of companies had one architect per project, with small numbers reporting more. However, 44\% of responses on this were inconclusive.

\noindent\textbf{Experience of Respondents:} Regarding experience levels, 44\% of the professionals had between one and five years of experience, 23\% had between six and ten years, and a small minority had over 11 years of experience. About 28\% of respondents indicated that this question was not applicable.

\noindent\textbf{Salaries of Software Architects and Non-Architects Performing Architect Duties:} In terms of salaries (converted into USD), 23\% of software architects earned between \$2,000 and \$4,000 per month, while 19\% earned between \$800 and \$1,600. A small percentage earned more than \$4,000, and about 46\% of respondents did not provide salary information. For those who perform architect duties without the formal title, 28\% earned between \$800 and \$1,600, 18\% earned more than \$2,000, and 35\% did not provide applicable data.

\vspace{-0.2cm}

\subsection{RQ3: Software Architects Practice}

The thematic analysis of responses revealed that software architects in Brazilian IT companies engage in a diverse range of activities, categorized into seven primary themes (Figure 1). The thematic analysis of interview responses indicates that software architects in Brazilian IT companies undertake a multifaceted range of activities, which can be grouped into seven primary themes. 

\textbf{Software development} emerged as the predominant activity (reported in 27 instances) among software architects, involving direct engagement in coding, module implementation, and contributing to codebases. This hands-on involvement ensures that architects maintain a deep technical understanding of the systems they design. Technically, this role demands proficiency in programming languages like Java, C\#, or Python, familiarity with development frameworks such as Spring Boot or .NET Core, and adherence to coding standards and best practices. Architects often participate in code reviews and contribute to continuous integration and deployment pipelines, utilizing agile methodologies for iterative development. This aligns with global trends where architects balance design and implementation, ensuring architectural decisions are grounded in practical feasibility \cite{babar2009exploratory}.

\textbf{Architecture design} is another significant role (reported in 25 instances), requiring a deep understanding of architectural patterns like microservices or event-driven architecture, and design principles such as SOLID or DRY. Architects employ modeling languages like UML to create diagrams representing system components and interactions, utilizing tools like Enterprise Architect.

\textbf{Defining technical solutions} that align with business requirements is also a key responsibility (reported in 20 instances). Architects analyze stakeholder needs, translate them into technical specifications, and design solutions that meet both functional and non-functional requirements. This involves requirements engineering and technology evaluation, assessing trade-offs between different technologies considering factors like performance, cost, and security. Studies highlight that architects' involvement in solution definition significantly reduces project risks associated with misalignment between technical deliverables and business goals \cite{Bass:2012}.

\textbf{Providing technical mentoring, and support} to development teams is another responsibility, albeit less frequently mentioned (reported in 15 instances). This involves facilitating knowledge transfer, conducting training sessions, and promoting best practices in coding and design. Architects often serve as technical consultants within teams, assisting with problem-solving and decision-making.

Architects are also responsible for \textbf{aligning technology strategies} with business objectives (reported in 14 instances). This involves evaluating emerging technologies like AI/ML and determining their applicability within the organizational context. Technically, this requires staying abreast of industry trends and assessing the impact of adopting new technologies on existing systems.

Some architects take on \textbf{project management} responsibilities (reported in 10 instances), overseeing timelines, resource allocation, and deliverables. This role bridges technical and managerial domains, requiring knowledge of project management methodologies like Scrum or PMBOK. Architects coordinate cross-functional teams, manage stakeholder expectations, and ensure that projects adhere to scope, budget, and quality standards. %The inclusion of architects in project management roles is supported by findings that technical expertise combined with management skills leads to better project outcomes \cite{jolley2021project}.

Lastly, architects play a role in \textbf{driving innovation} within organizations (reported in 7 occurrences). They research and experiment with cutting-edge technologies, prototype new solutions, and promote a culture of continuous improvement. Technically, this may involve developing proof-of-concepts using emerging technologies like serverless computing or exploring new architectural paradigms. Global studies indicate that architects are pivotal in fostering innovation, with organizations that empower architects to explore new technologies reporting innovation performance \cite{Garlan:2010}. 
\vspace{-0.2cm}
\section{Discussion}
\label{discussion}

This study provides an examination of software architecture practices within Brazilian companies, revealing significant disparities in the formal recognition of software architects, their responsibilities, and compensation structures. These discrepancies highlight systemic challenges in aligning software architecture roles with organizational objectives and the rapidly evolving demands of the global market. The informal delegation of architectural responsibilities to non-designated professionals can pose risks to the integrity, scalability, and maintainability of software systems.

%\section{Comparative Sectoral Analysis (RQ1)}

About the \textbf{companies' characteristics (RQ1)}, we could notice from the data that the majority of the companies are relatively mature (37\% are between 11 and 30 years old). A significant portion (26\%) are quite young (under 10 years), and there’s a notable outlier: a company operating for over 500 years, suggesting the sample includes at least one extremely historic institution (a bank). Small companies (serving 1–10 customers) are fairly common (22\%), but there’s a balanced representation of medium-sized companies (16\% each for 11–100 and 101–15,000 customers). A notable chunk (17\%) work for very large companies with over 1 million customers.

Most companies (44\%) manage a small number of parallel projects (up to 10), indicating relatively focused project management. A portion (24\%) manage between 11–50 projects. A small minority (9\%) manage over 1,000 projects, suggesting the presence of very large/complex organizations. The 11\% inconclusive data suggests some challenges in reporting how projects are defined.

We could observe that public-sector organizations in Brazil exhibit a more balanced distribution of technical and collaborative responsibilities. Globally, government IT projects often emphasize collaboration and knowledge transfer to ensure that public services remain reliable and accessible. This approach aligns with observations in other countries where public-sector organizations prioritize stability and compliance, ensuring the sustainability of their systems over the long term \cite{united2020government}.

When regarding to the \textbf{role and presence of architects (RQ2)}, the data suggest that the software architecting in Brazil is frequently performed informally, with many respondents showing a lack of clarity about the number of architects or their assignment to projects. The practice is common, but the formalization of the role remains limited. Research in software engineering underscores the importance of formalizing the software architect role to ensure systems are designed with scalability, maintainability, and security in mind. A study by \cite{Kruchten:Past:Present:Future:Architecture} highlights that organizations neglecting formal architecture roles often encounter increased costs due to rework, system refactoring, and unanticipated maintenance efforts. For Brazilian companies aiming to enhance competitiveness and deliver high-quality software solutions, recognizing and supporting the software architect role is a strategic necessity.

 When it comes to the \textbf{software architects practice (RQ3)}, we also observed from the data that the environment has led to a strong emphasis on core technical activities like software development and architectural design, which can be important for ensure technical consistency, but also worrying, since there is some overlapping in the roles. This also demands more investigation.

We also triangulated the data to obtain other insights. From that, we gather additional findings. A total of 43 companies reported having architects in their teams, while 21 said that they have people that informally perform the role of software architect. This is a bit worrying, since it means that among the companies that reported having software architects, 48.84\% also indicated that there is someone within the company who performs the role of a software architect informally, without being formally hired under that title. This leads to the need of further investigation to understand why this observed phenomenon happens.

When analyzing whether the companies with software architects mostly are those where software development is a core activity, we found that 63 companies indicated that software development is a core activity. Among the companies where software development is a core activity, 32 companies have software architects hired specifically to fill that role. This indicates that 27\% of the total companies surveyed, which have software architects, are primarily engaged in software development as their core activity. And more, this shows that about a half of the companies whose main activity is software development does not have a software architect formally hired as such, which is also worrying. And something even more alarming: Among the companies where software development is a core activity and have software architects hired specifically to fill that role, only 2 companies are classified as small-sized; which means that medium-sized and large companies are not valuing having software architects in their team. However, we observed that companies where software development is a core activity tend to have more software architects (3.5 on average) compared to those where software development is a secondary activity or for internal consumption (about 2 architects as average).

We also investigated the average monthly salary ranges for software architects in the respective regions. Upon analyzing the survey data, it was found that the highest salaries for software architects in each region of Brazil were as follows: in the North Region, Amazonas (AM) offered the highest salary at R\$ 20,000.00; in the Northeast Region, Ceará (CE) offered the highest salary at R\$ 20,000.00; in the Central-West Region, Goiás (GO) offered the highest salary at R\$ 20,000.00; in the Southeast Region, São Paulo (SP) offers the highest salary at R\$ 20,000.00; and in the South Region, Santa Catarina (SC) offered the highest salary at R\$ 20,000.00. Upon reviewing the survey data, no company offered a salary above R\$ 20,000.00 for software architects at the time of the survey conduction. 

We also analyzed if there was a significant difference in the activities performed by architects in different states. We found the following data, considering only some states due to space restrictions: 
São Paulo (10 unique activities), Rio de Janeiro (8 unique activities)
Goiás (7 unique activities), Santa Catarina (6 unique activities) and
Amazonas (5 unique activities). This variation indicates that the professional roles of software architects do differ significantly between states, with São Paulo having the most diverse set of activities being performed.

Upon analyzing the survey data, it was found that the maximum monthly salary for software architects in multinational companies is R\$ 20,000.00, while the maximum monthly salary for software architects in national companies is also R\$ 20,000.00. This indicates that both multinational and national companies offer similar maximum salary ranges for software architects.

\subsection{Threats to Validity and Limitations}
\label{threats}

\noindent \textbf{Time Frame.} %Software architecture evolves rapidly with technological advancements. 
Our study, conducted during COVID-19 (2020–2021), may not fully reflect current realities, particularly regarding architects’ roles and compensation, which may have shifted due to economic changes. While the tech sector has seen major shifts, survey results capture a specific period. Despite this, our large sample and national scope provide valuable insights, especially given the lack of prior studies in Brazil. 
% \noindent \textbf{Population.} We surveyed 105 Brazilian companies, ensuring broad national representation across 24 of Brazil’s 27 states. Compared to studies with smaller samples (e.g., 72, 32, and 100 respondents), ours offers a more complete view of the software industry. While sample size limitations exist, our extensive coverage strengthens the validity of our findings.  \\

% \noindent \textbf{Qualitative Analysis.} Our thematic analysis has limitations, including potential researcher bias in theme identification and incomplete data saturation. Participant bias may also influence responses. To address these issues, one researcher conducted initial coding, with others reviewing findings to improve reliability.

%\noindent \textbf{Time Frame}. Software architecture evolves rapidly. Conducted during COVID-19 (2020–2021), our study may not fully reflect current realities, particularly in architects’ roles and compensation. While the tech sector has changed, our large sample and national scope provide valuable insights, especially given the lack of prior studies in Brazil.  
\noindent \textbf{Population.} We surveyed 105 Brazilian companies across 24 states, offering broad national representation. Compared to smaller studies (e.g., 72, 32, and 100 respondents), ours presents a more comprehensive industry view. While sample size limitations exist, our wide coverage enhances validity.  

\noindent \textbf{Qualitative Analysis.} Our thematic analysis has limitations, such as researcher bias in theme identification and incomplete data saturation. Participant bias may also impact responses. To mitigate this, one researcher conducted initial coding, with others reviewing findings for reliability.

\section{Final Remarks}
\label{final}

Our study reveals critical deficiencies in the recognition and remuneration of software architects in Brazilian companies, with notable regional disparities. Assigning architectural tasks to non-specialists can harm software quality, though firms directly engaged in development align better with global standards, suggesting potential industry improvements.  

To remain competitive, Brazilian companies must adopt global architectural practices, integrating AI-driven design, cloud-native solutions, and automated compliance via Infrastructure as Code (IaC). These innovations enhance efficiency but demand a redefined architect role focused on adaptability. Future studies should explore technology integration and compare Brazil with markets like India and Chile for best practices. Addressing these gaps and embracing technology can boost Brazil’s software industry, improving project outcomes and economic impact.

\section*{Artifacts Availability}

Data and high resolution figures are available at \url{https://doi.org/10.5281/zenodo.15252127}.

\bibliographystyle{ACM-Reference-Format} 
\bibliography{bib}

\end{document}